%Paper: hep-ph/9303237
%From: DELDUCA@SLACVM.SLAC.Stanford.EDU
%Date: Tue, 09 Mar 1993   14:56 -0800 (PST)

% A hard copy of the text and the figures is available upon request
% from the SLAC Publications Dept.  CRYSTAL at SLACVM

%macropackage=phyzzx

% these macros are to be used in math mode:
\def\Re{{\rm Re}}
\def\Im{{\rm Im}}
\def\half{{1\over 2}}

% these macros are to be used in text mode:
%\def\brk{\hfill\break}
%%%%%%%%%%%%%%%%%
%\draft
\overfullrule=0pt

\pubnum{6065}
\date{February, 1993}
 \pubtype{T/E}
 \titlepage
 \title{%
Computation of Mini-Jet Inclusive Cross Sections\doeack}
\author{Vittorio Del Duca, Michael E. Peskin, and Wai-Keung Tang}
\SLAC
\abstract{We apply the theory of parton-parton total cross sections
at large $s$, due to Lipatov and collaborators, to compute the
inclusive cross section for jets which accompany a large $s$
parton scattering process.}
\submit{Physics Letters B}
\endpage
\vfill

  The enormous dynamic range for QCD processes opened by the
Tevatron collider has given a new impetus to detailed investigations
of the dynamics of quark and gluon scattering.  Though the most
prominent experimental investigations have involved the cross section
for 2-jet production, there are also new results on 3- and multi-jet
processes.  In these higher-order processes, there are two or more
different Lorentz invariants which
set the scale of the  momentum transfer, and, typically, these
invariants may differ by large factors.  The study of QCD processes
with large ratios of invariants---{\it semihard\/} processes---is
complicated theoretically because it typically involves the resummation
of infinite classes of Feynman diagrams.  It is complicated
experimentally because it requires jet detection in a large rapidity
interval.   Nevertheless, this regime is a fascinating one, with new and
nontrivial applications of QCD perturbation theory.   In addition,
as Bjorken has recently emphasized, the study of jet correlations over
large rapidity intervals may offer interesting signatures of new
physics.\Ref\BJ{J. D. Bjorken, \sl Phys. Rev. \bf D47 \rm (1992) 101.}

\REF\BFKLone{L. N. Lipatov, \sl Sov. J. Nucl. Phys. \bf 23 \rm
(1976) 338.}
\REF\BFKLtwo{
E. A. Kuraev, L. N. Lipatov,
and V. S. Fadin, \sl Sov. Phys. JETP \bf 44 \rm (1976) 443, \bf
 45 \rm (1977) 199.}
\REF\BFKLthree{Ya. Ya.
Balitsky and L. N. Lipatov, \sl Sov. J. Nucl. Phys.
\bf 28 \rm (1978) 822; \brk
L. N. Lipatov, \sl Sov. Phys. JETP \bf 63 \rm (1986) 904.}
   In this paper, we will discuss the following situation, which
involves jet dynamics with a large ratio of invariants.   Consider
a process in which gluons or quarks scatter with large center-of-mass
energy and only moderately large transverse momentum.   Call the squared
center-of-mass energy of the original parton-parton scattering $s$,
and let $m$ be a typical value of the final parton transverse momenta.
In the semihard regime, $s >> m^2 >> \Lambda_{QCD}^2$,  Lipatov and
collaborators\refmark{\BFKLone-\BFKLthree}
 have shown that, in this regime, the rapidity interval
between the scattered partons is filled in by the radiation of additional
gluons, roughly uniformly spaced in rapidity, all with transverse
momenta of order $m$.   In principle, the properties of these
radiated gluons can be observed by choosing events with two jets at
large positive and negative rapidity and then measuring the production
rate for jets at intermediate rapidities which accompany these
jets.  In this paper, we will call the jets at large rapidity
{\it tagging jets\/}, and we will call the jets that accompany them
{\it minijets\/}.
 A few results on the spectrum of these accompanying
gluon jets are known from the work of
Balitsky, Fadin, Kuraev, and Lipatov
(BFKL),
refs. \BFKLone--\BFKLthree,
and from succeeding work of Levin and Ryskin.\Ref\LR{M. G.  Ryskin,
    \sl Sov. J. Nucl. Phys. \bf 32 \rm (1980) 133; \brk
  E. M. Levin and M. G. Ryskin, \sl Sov. J. Nucl. Phys.
            \bf 32 \rm (1980) 413.}
 In this paper, we return to this question and give a systematic
procedure for computing the inclusive spectrum of accompanying jets,
along with numerical  estimates for  Tevatron  and SSC energies.

\noindent{\bf Total Cross Section for Tagging Jets}

 In order to understand the
    spectrum of minijets which accompany a set of tagging jets, we should
    first review the QCD prediction for the total cross section
 for producing these tagging
jets.  This prediction has been presented by Mueller and
Navelet,\Ref\MN{A. H. Mueller and H. Navelet, \sl Nucl. Phys.
            \bf   B282 \rm (1987) 727.}
using the expression for the asymptotic
parton-parton total cross section  which
is the principal result of the BFKL analysis.
We will use the ingredients in this prediction to construct the
related inclusive cross sections.

  Following Ref. \MN, we consider the scattering of two hadrons of
momenta $p_A$ and $p_B$ in the center-of-mass frame, with the $z$ axis
along the beam momenta, and we imagine that we
 tag two jets at the extremes of
the Lego plot, with the rapidity
interval between them filled with minijets.  The tagging jets
can be characterized by their transverse momenta
and by their longitudinal fractions $x_A$, $x_B$ with
respect to their parent hadrons.  It is simplest to consider the
cross section for producing two tagging jets with transverse momenta
greater than a minimum value $m$.  Then
$$\eqalign{
 {d\sigma \over dx_A dx_B}&(AB\rightarrow j(x_A) j(x_B) + X) \cr
&  = \prod_{i=A,B}
   \biggl[G(x_i,m^2) + 4/9 \sum_f [Q_f(x_i,m^2) +
   \bar Q_f(x_i,m^2)]\biggr]  \cdot
   \sigma_{tot}(s),  \cr}
\eqn\totaltagrate$$
where $s = 2p_A\cdot p_B x_A x_B$ is the parton-parton squared
center-of-mass energy, and
 $\sigma_{tot}$ is the BFKL total cross section for gluon-gluon
scattering, which we will discuss in a moment.   Eq. \totaltagrate\
includes the effects of quarks using the observation of Combridge
and Maxwell that, in a process with large rapidity intervals, the
leading contribution to any scattering process comes from gluon
exchange in the crossed channel.\Ref\CM {B. L. Combridge
and C. J. Maxwell,
\sl Nucl. Phys. \bf B239 \rm (1984) 429.}
 In writing \totaltagrate, we assume that
the values of $x_A$, $x_B$ are  sufficiently large that the parton
distributions can be computed from ordinary Altarelli-Parisi evolution;
semi-hard QCD adds additional complications when these fractions
become small.\Ref\smallpart{J. C. Collins and R. K. Ellis, \sl
          Nucl. Phys. \bf B360 \rm (1991) 3; \brk S. Catani, M.
             Ciafaloni, and F. Hautmann, \sl Nucl. Phys. \bf B366
               \rm (1991) 135.}

The core of eq. \totaltagrate\ is the BFKL function $\sigma_{tot}(s)$,
which is given by
$$ \sigma_{tot}(s) = {8 \over N_c^2 - 1} {\pi N_c^2 \alpha_s^2 \over
           2 m^2}
   F(Y), \eqn\sigmatot$$
where $N_c = 3$ is the number of colors in QCD,
 $Y = \log(s/m^2)$, and $F(Y)$ is a dimensionless function
which we will discuss below.  The strong coupling constant is evaluated
at a scale $m^2$; the running of $\alpha_s$ is subleading in the BFKL
theory.
 Setting $F = 1$ gives the cross section
for gluon-gluon scattering at the lowest order of QCD, integrated
over transverse momenta with $|k_{i\perp}| \geq m$, in the limit
$s>>m^2$.

  \FIG\BFKLg{BFKL resummation: (a) emission of gluons; (b) construction
    of the octet amplitude; (c) construction of the singlet amplitude.}

The BFKL theory systematically corrects the lowest-order QCD result
for $\sigma_{tot}$ by summing the leading logarithms of $(s/m^2)$.
This is done in three stages, as shown in Fig. \BFKLg.
First, one simplifies the lowest-order QCD diagrams for multigluon
production, shown in Fig. \BFKLg(a), for the case in which the emitted
gluons are widely separated in rapidity. The gluon emission vertex is
replaced by a non-local gauge-invariant effective vertex
\refmark\BFKLone. Next, one sums the leading
corrections to the forward
amplitude with  color octet in the $t$-channel, as
shown in Fig.  \BFKLg(b).  The result has the form of a Regge pole
with an infrared-sensitive trajectory.\refmark\BFKLtwo
Finally, one uses this resummed, effective
gluon exchange to compute the forward amplitude with color singlet
in the $t$-channel.\refmark\BFKLthree
  The imaginary part of this object is the
gluon-gluon total cross section.

  The last two step of this procedure involve the solution of
integral equations constructed by BFKL.   In the final step,
 the integral equation can be solved explicitly for the imaginary
part of the forward amplitude to find
tagging jets with transverse momenta $k_{A\perp}$, $k_{B\perp}$,
both greater than $m$, separated by a rapidity interval $Y$.
  The solution of the equation involves a
Laplace transform with respect to rapidity.  Then the amplitude is
given by
$$ f(k_{A\perp}, k_{B\perp}, Y) =
   \int_{a-i\infty}^{a+i\infty} {d\omega \over 2\pi i} \
   e^{\omega Y} \ f_{\omega}(k_{A\perp}, k_{B\perp}),
\eqn\fform$$
where the Laplace transform has the representation
$$ f_{\omega}(k_{A\perp}, k_{B\perp}) = {1\over (2\pi)^2}
   \sum_n e^{in(\phi_A - \phi_B)} \int d\nu
   {(k_{A\perp}^2)^{-1/2 +i\nu} (k_{B\perp}^2)^{-1/2 -i\nu}
   \over \omega -\omega(n,\nu)}.
\eqn\Lapfform$$
In this equation,
$\phi_A - \phi_B$ is the azimuthal angle between the transverse
momenta of the tagged jets, and
$$ \omega(n,\nu) = {2N_c \alpha_s \over\pi}\bigl[ \psi(1) - \Re\psi
   ({|n|+1\over 2} +i\nu) \bigr],
\eqn\omegeval$$
with $\psi(z)$ the standard logarithmic derivative of the Gamma function.
Throughout this paper, we ignore azimuthal correlations and keep only
the leading, $n=0$, term of \Lapfform.  Near $\nu = 0$, $\omega(\nu) =
\omega(0,\nu)$ has the expansion
$$   \omega(\nu)  =  A  -  B\nu^2 + \cdots ,
\eqn\omegaexpand$$
with
$$     A =  {4N_c\alpha_s\over \pi}\log 2
   , \qquad    B = {14N_c\alpha_s\over \pi} \zeta(3)
\eqn\AandBforms$$

The integral over $\omega$ in \fform\ can be done easily by picking up
the pole.  Then we may integrate over
 the jet transverse momenta to obtain
the enhancement factor  $F$ in eq. \sigmatot.   The integrals over
$k_{A\perp}$ and $k_{B\perp}$ are singular and depend on the cutoff $m$;
this gives the factor $m^{-2}$ in  \sigmatot.  Comparing with that
formula more closely, we find
$$ F(Y) =  \int_{-\infty}^{\infty}
   {d\nu\over 2 \pi} {1 \over \nu^2 + 1/4} e^{Y\omega(\nu)}.
\eqn\ffinal$$
 The exponential growth of
$F(Y)$ with the rapidity interval is associated with minijet
production.   Using \omegaexpand\ to expand about the saddle point
at $\nu = 0$, we can see that $F(Y)$
 has the asymptotic behavior
$$    F(Y) \sim {e^{(4\log 2) z}\over \sqrt{7\zeta(3) \pi z/2} },
\qquad {\rm with }\   z = {N_c\alpha_s\over \pi} Y .
\eqn\asymptotsZ$$
Mueller and Navelet showed that this asymptotic form is
an accurate representation for $z> 0.2$.

\noindent{\bf Mini-jet inclusive cross section}

 The BFKL total cross section is a sum over multi-jet emission
processes.  Thus, it is not difficult to pull this amplitude apart
and find the contributions from final states with a jet in a
specific region of phase space.  The cross section for producing
   \FIG\InclSum{Sum of graphs leading to the inclusive jet
                  cross section.}
a jet at rapidity $y$ and transverse momentum $q_\perp$
in association with the tagging jets described in the previous
section may be computed as the sum of diagrams shown in Fig. \InclSum.
We use the BFKL propagator  \fform\ to represent the ladder and
Lipatov's vertex from ref. \BFKLone\ to represent the gluon emission.
For the moment, we retain the dependence on the transverse momenta
$k_{A\perp}$, $k_{B\perp}$ of the tagging jets.
Then the inclusive cross section for minijet emission is given by
$$ \eqalign{
 {1\over \sigma_{tot}}{ d\sigma\over d y d^2 q_\perp
           d^2 k_{A\perp} d^2 k_{B\perp}}
= & {16 \alpha_s N_c\over \pi} { m^2\over F(Y)}
\int{ d^2k_{1\perp}\over (2\pi)^2}
{ d^2 k_{2\perp}\over (2\pi)^2}  (2\pi)^2\delta^{(2)}(k_{1\perp} -
k_{2\perp} -  q_{\perp} )  \cr
   &\quad \cdot   {1\over k_{A\perp}^2
       k_{B\perp}^2  q_\perp^2}  f(k_{A\perp},k_{1\perp},y_A-
          y) f(k_{2\perp}, k_{B\perp}, y - y_B) , \cr }
\eqn\oneincleq$$
where $y_A$, $y_B$ are the rapidities of the tagging jets, with
$Y = y_A - y_B$.
This formula has been derived previously by  Levin and
Ryskin.\refmark{\LR}

  As before, we ignore all correlations in azimuthal
angle by integrating or averaging each transverse momentum over
its angle $\phi$.
 Note that the correlations in $\phi$ are, in
any event, subleading.
With this simplification, we can represent the momentum delta function
in \oneincleq\ as
$$ (2\pi)^2  \delta^{(2)}(k_{1\perp} -   k_{2\perp} -  q_{\perp} )  =
   \int d^2 b  e^{ib\cdot(k_{1\perp} - k_{2\perp} - q_\perp)}
   = \int d^2b\,  J_0(bk_{1\perp}) J_0(bk_{2\perp}) J_0(bq_\perp).
\eqn\deltareduce$$
By introducing the representation \Lapfform, we can now perform the
integrals over the four $k_{i\perp}$. Notice that the integrals over
$k_{A\perp}$, $k_{B\perp}$ depend on the infrared cutoff, as in the
derivation of \ffinal.  However, the integrals over $k_{1\perp}$,
$k_{2\perp}$ converge in the infrared.
  The resulting expression is a
function of two Laplace transform variables, which we will call
$\nu_1$, $\nu_2$.  We are left with the integral over $b$, which
reduces to
$$  \int^\infty_0 db\,  b^{-1 +2 i (\nu_1 - \nu_2)}  J_0(bq_\perp)
\eqn\finalbint$$
To make this integral well-defined, we deform the contours of
integration over the $\nu_i$ so that $\Im(\nu_2) > \Im(\nu_1)$.
We must keep  $|\Im (\nu_i)| < \half$ in order that the integrals
over the $k_{i\perp}$ converge.
We will show below that there is a saddle point  for the $\nu_i$
with this property.

  After performing the integral over $b$, we are left with the
following expression for the inclusive
cross section for the production of
one minijet with rapidity $y$ and transverse momentum  $q$
in association with tagging jets with transverse momentum greater
than $m$:
$$ \eqalign{
 {1\over \sigma_{tot}}{ d\sigma\over d y d^2 q_\perp }
= & { 1\over F(Y)}  \int^\infty_{-\infty}\!{ d\nu_1\over 2\pi}{ d\nu_2
   \over 2\pi}
e^{\omega(\nu_1)(y_A-y)}
          e^{\omega(\nu_2)(y-y_B)} {N_c\alpha_s\over \pi^2 q^2_\perp}
    \bigl({m^2\over q_\perp^2}\bigr)^{i(\nu_1-\nu_2)}
          \Bigl({-i\over \nu_1 - \nu_2}\Bigr)  \cr
&\qquad \biggl[{1\over (\half - i\nu_1)}{\Gamma(\half - i \nu_1) \over
  \Gamma(\half + i \nu_1)}{\Gamma(1 + i(\nu_1-\nu_2))\over
                        \Gamma(1 - i(\nu_1-\nu_2)) }
{\Gamma(\half + i \nu_2) \over  \Gamma(\half -i \nu_2)}
          {1\over (\half + i\nu_2)}\biggr]. \cr}
\eqn\oneinclgams$$
The variables $\nu_i$ are integrated along contours with
$  \half > \Im (\nu_2) > \Im (\nu_1) > - \half$.

 The same method yields an integral expression for the $n$-minijet
inclusive cross section.  The integration variables are the $n+1$
Laplace transform variables of the intermediate
BFKL propagators. By the
argument given above, these are ordered along the imaginary  axis
from the tagging jet $B$ to the tagging jet $A$:  $\half > \Im(\nu_{n+1})
 > \cdots > \Im (\nu_1) > -\half$.  The full inclusive cross section
is given by
$$ \eqalign{{1\over \sigma_{tot}}&
 {d\sigma \over  \prod_{i=1}^n
   d^2q_{i\perp} dy_i} \cr
   & = {1\over F(Y)}
   \prod_{i=1}^{n+1}
   \int {d\nu_i \over 2\pi}
   e^{\omega(\nu_i)(y_{i-1}-y_i)}
 \cdot\prod_{i=1}^{n} \bigg\{{N_c\alpha_s\over \pi^2
 q_{i\perp}^2}
  \Bigl( {q_{i\perp}^2 \over m^2}
   \Bigr)^{-i(\nu_i - \nu_{i+1})}
   \Bigl({-i \over \nu_i - \nu_{i+1}}\Bigr) \bigg\}
\cr  &\qquad \cdot\biggl[
    {1 \over (\half-i\nu_1)}
   {\Gamma(\half-i\nu_1) \over \Gamma(\half+i\nu_1)}
   \bigg\{ \prod_{i=1}^n
   {\Gamma[1+i(\nu_i-\nu_{i+1})] \over
   \Gamma[1-i(\nu_i-\nu_{i+1})]}
   \bigg\} {\Gamma(\half+i\nu_{n+1}) \over
   \Gamma(\half-i\nu_{n+1})}
   {1 \over (\half+i\nu_{n+1})}\biggr].\cr} \eqn\seven$$

\noindent{\bf Asymptotic Evaluation}

   In eq. \asymptotsZ, we simplified the integral formula for the
total jet-jet cross section by evaluating the integral asymptotically
around an appropriate saddle point.  It is straightforward to
evaluate the one-jet inclusive cross section asymptotically in the
same way.  We will find a joint saddle point in the variables $\nu_1$,
$\nu_2$ and expand the double integral around this point.  The expansion
will be valid in the limit in which $(y_A  - y)$ and
 $(y-y_B)$ are both large. We neglect the terms in the square
brackets in eq. \seven, since at asymptotic energies the saddle point
will be very close to the origin in the plane $\nu_1$, $\nu_2$.

  The joint saddle point will occur with both of the $\nu_i$ on the
imaginary axis: $\nu_i = - i n_i$. Using the expansion  \omegaexpand,
we find the saddle point conditions:
$$ \eqalign{
  2(y_A-y) B n_1  - \log(q_\perp^2/m^2)  - {1\over n_1-n_2} &= 0 \cr
  2(y-y_B) B n_2  + \log(q_\perp^2/m^2)  + {1\over n_1-n_2} &= 0. \cr }
\eqn\saddletwo$$
{}From these equations,
$$    n_1 = (y - y_B) \eta;\qquad   n_2 = -(y_A-y) \eta,
\eqn\findthens$$
with
$$  \eta = {\pi\log(q^2_\perp/m^2)\over 56N_c \zeta(3)  \alpha_s (y_A-y)
                  (y-y_B) } \bigl[ 1 + a ],
\eqn\valofetay$$
and
$$ a = \sqrt{1+{112N_c\zeta(3)\alpha_s(y_A-y)(y-y_B)
    \over \pi Y\ln^2(q_{\perp}^2/m^2)}}.
\eqn\valofa$$
Expanding about this saddle point, we find
$$ {1\over \sigma_{tot}}
{d\sigma \over dy  d\ln q_{\perp}^2} =
   {N_c\alpha_s \over 2\pi^{5/2}}
   \sqrt{1-{1\over a}} exp\left[ -{1\over a-1} + 1/2 \right],
   \eqn\eight$$
with $a$ as in \valofa.

  According to this equation, the jet inclusive cross section, falls
off faster than the scale invariant dependence $(d^2 q_\perp/q_\perp^2) $
 expected
for an approximately fixed value of $\alpha_s$.  At small transverse
momentum, where we can ignore the $\log(q^2_\perp/m^2)$ terms in
solving \saddletwo,
there only is a small modification  by a factor $(q^2_\perp)^{-Y\eta}$.
At large transverse momentum, the cross section falls off faster than
any power of $q_\perp$.  This latter, doubly asymptotic, limit has
been found previously by Ryskin using another method.\refmark\LR

\noindent{\bf Numerical Evaluation}

The formula \oneinclgams\ is also an appropriate starting point for a
numerical evaluation of the one jet inclusive cross section.
It is important to study the exact behavior of this formula
numerically, for two reasons.  First, the total rapidity interval
for jet-jet scattering processes is not large at energies now
available, and so it might be troublesome to  divide this interval
into several pieces, each of which must be large to justify an
asymptotic analysis.  Second, this problem is exacerbated by the
presence of the poles in \oneinclgams\ which restrict the
region of the imaginary axis through which the $\nu_i$ contours
can pass.  At asymptotic energies, the saddle points found in the
previous section lie very close to $\nu_i = 0$, and so these
poles are not relevant.  However, the
evaluation of the saddle point locations  with realistic parameters
puts  $n_1$ and $n_2$ close to $\pm\half$. Thus, we might expect large
discrepancies for the saddle point result in realistic cases, and,
indeed, we will find this below.

    To compute the minijet cross section explicitly, we evaluated
the double integral in \oneinclgams \ numerically using the
contours $\Im\nu_1, \Im\nu_2 = \pm {1\over 4}$.  We evaluated
$\alpha_s = \alpha_s(q_\perp)$, scaled from $\alpha_s(m_Z) = 0.12$
using 1-loop evolution with 5 flavors.

   \FIG\Compareasy{Comparison of the exact and asymptotic evaluations of
   the one jet inclusive cross section as a function of the jet transverse
   momentum, at the parton center-of-mass energies
   $\sqrt{s}= 1, 10^3, 10^6, 10^9$ TeV, and $m = 20$ GeV.
   The upper curves represent the exact evaluation of
   integral \oneinclgams, and the lower curves its asymptotic
   evaluation \eight.}
For \ffinal, we saw that the saddle point approximation was
an accurate one.  However, this turns out not to be true for the
asymptotic evaluation of \oneinclgams.  In Fig. \Compareasy,
we compare the exact and asymptotic formulae for the one jet
inclusive cross section for  gluon-gluon scattering
processes at center of mass energies from 1  to 10$^{9}$ TeV,
with  $m= 20$ GeV.
There are large discrepancies between the two formulae which
disappear only extremely slowly, as the inverse of the logarithm
of the energy.  In fact, the asymptotic formula is never accurate.
Since the total cross section
\asymptotsZ\ rises as a power of $s$, the contribution from
the exchange of one gluon ladder must at some point be  unitarized
by contributions from multiple ladder
exchanges.\Ref\GRL{L. V. Gribov, E. M. Levin, and M. G. Ryskin,
\sl Phys. Repts. \bf 100 \rm (1983) 1.}
This correction is already important at 100 TeV, and changes the
one jet inclusive cross section essentially at higher energies.

  However, even if the asymptotic formula is not valid, we can
estimate the minijet inclusive cross section directly from
\oneinclgams.
It is important to recall that this equation, as well as \ffinal,
 involve additional asymptotic
approximations.  The BFKL integral equation is derived under the
assumption that the emitted gluons are widely separated in rapidity.
The solution of the color octet exchange problem shown in Fig. 1(b)
involves the assumption that the leading Regge
pole in the $\omega$ plane dominates over
the remaining singularities.  However, there is no reason why either of
these effects should lead to large corrections under realistic
conditions.

    \FIG\TevandSSC{One jet inclusive cross section as a function of
         the transverse momentum of the jet for typical Tevatron
            and SSC energies. We choose $\sqrt{s} = 0.18$ and $4$ TeV
            respectively for the Tevatron and the SSC parton
            center-of-mass energies, and we label the curves according
            to the minimum jet transverse momentum.}
  With this set of approximations, then, we present in Fig. \TevandSSC\
predictions from \oneinclgams\ for the minijet inclusive cross section at
Tevatron and SSC energies, taking the longitudinal fractions $x_A$, $x_B$
of the
tagging jets momenta to be $0.1$.  The cross section is computed for an
observed jet in the center of the rapidity interval between the
tagging jets.  However, the minijet
cross section is almost independent of
the position  of the jet in the rapidity interval.  These curves, and
also those of Fig. \Compareasy, show a shift of the transverse
momentum distribution toward higher values as the energy of the
original scattering process---and, therefore, the length of the
gluon cascade---increases.  This effect can also be seen in eq. \eight,
although the dependence of that formula on $q_\perp$ and $y$
does not match our quantitative results.

   The final situation in which we are
   left is somewhat ambiguous.  On
one hand, we have improved the  understanding of the physics
of the  BFKL total cross section.  We have presented predictions from
the BFKL theory for the one minijet
 inclusive cross section, and we have
presented a formula which can be used to evaluate higher order
jet correlations.  However, we have found that the natural asymptotic
evaluations of these formulae are not accurate at the energies of
present colliders.  It will be interesting to see whether the
estimates presented in Fig. \TevandSSC\ will be confirmed
experimentally.  However, we did not succeed in presenting
 a quantitative
and characteristic prediction which can be used to confirm the
 BFKL resummation.
That remains an interesting problem for the future.

\bigskip

\noindent{\bf Acknowledgements}

We are very grateful to bj. Bjorken and Al Mueller for encouraging
us to investigate this problem.  We also thank
 Lev Lipatov and Eugene Levin
for useful discussions.

\endpage

\refout
\endpage

\figout

\end
\bye